\documentclass[twocolumn,showpacs,prd]{revtex4}
\usepackage{mathrsfs}
\usepackage{amsmath, txfonts}
\usepackage{graphicx,,booktabs,bm}
\usepackage{longtable,lscape}
\usepackage{overpic,multirow}
\usepackage{color}
\usepackage{amssymb}
\def\half{{\textstyle{1\over 2}}}
\def\thalf{{\textstyle{3\over 2}}}

\usepackage{indentfirst}
\begin{document}
\title{$Z_c(3900)$ as a resonance from the $D\bar{D}^*$ interaction}
\author{Jun He}\email{junhe@impcas.ac.cn}
\affiliation{
Theoretical Physics Division, Institute of Modern Physics, Chinese Academy of Sciences,Lanzhou~730000,China
}

\affiliation{
Research Center for Hadron and CSR Physics,
Lanzhou University and Institute of Modern Physics of CAS, Lanzhou 730000, China
}

\affiliation{
State Key Laboratory of Theoretical Physics, Institute of
Theoretical Physics, Chinese Academy of Sciences, Beijing  100190,China
}

\begin{abstract}

In this paper it is proposed that the charged charmonium-like state
$Z_c(3900)$ is a resonance above the threshold from the $D\bar{D}^*$
interaction. The $D\bar{D}^*$ interaction is described by
the one-boson exchange model with  light meson exchanges plus a short-range
$J/\psi$ exchange.  The scattering amplitude is calculated within a
Bethe-Salpeter equation approach and the poles near the $D\bar{D}^*$
threshold  are searched. In the isoscalar sector, two poles found under the
$D\bar{D}^*$ threshold, i.e., bound states, have the quantum numbers $I^G(J^{PC})=0^-(1^{+-})$ and
$0^+(1^{++})$. The latter can be related to the $X(3872)$. In the
isovector sector, a bound state with $I^G(J^P)=1^+(1^+)$ is found  with a large cutoff at about 3 GeV.  If a cutoff at about 2 GeV is
adopted with which a pole carrying the quantum number of the $X(3872)$  is
produced at an energy of about 3871 MeV, the pole for the bound state with
$1^+(1^+)$ runs across the threshold to a second Rienman sheet and becomes a resonance above
the $D\bar{D}^*$ threshold, which can be identified with the $Z_c(3900)$. With such a cutoff, the $D\bar{D}^{*}$ invariant mass spectrum is also investigated and the experimental
results found by BESIII can be reproduced.

\end{abstract}

\maketitle
\maketitle

\section{Introduction}

In the past decade, an amount of the so-called XYZ particles were found
in the facilities around the world, such as Belle, $BABAR$ and BESIII.  It is difficult to
put these particles into the conventional quark model frame, which
attracts physicists' special attention. An interesting observation is
that many of the XYZ  particles are near the threshold of two charmed
mesons. For example, the mass gap between the $X(3872)$ observed by
the Belle Collaboration \cite{Choi:2003ue}  and the $D^0\bar{D}^{*0}$
threshold is smaller than 1 MeV. The structure $Z_c(3900)$ observed by
 BESIII \cite{Ablikim:2013mio} is also very close to
the $D\bar{D}^*$ threshold.  It inspired the idea that these
particles originate  from the interaction of two mesons, such as
the hadronic molecular state and threshold effect.


In the literature, many efforts have been made to study the possibility of
interpreting the $X(3872)$ and the $Z_c(3900)$ as a hadronic molecular stat (that is, a bound state below the threshold),
such as calculations with the QCD sum rule
 \cite{Wang:2013daa,Zhang:2013aoa,Cui:2013yva}.  In
Refs.~\cite{Sun:2011uh,Sun:2012zzd}, the $B\bar{B}^*/D\bar{D}^*$
system was studied with a nonrelativistic one-boson exchange (OBE)
model by solving the Schr\"odinger equation. There does not exist a bound state
from the $D\bar{D}^*$ interaction, which can be identified with the $Z_c(3900)$ observed by BESIII. In Ref.~\cite{He:2014nya}, the $D\bar{D}^*$ interaction is
studied in a Bethe-Salpeter equation approach. A state with quantum
number $I^G (J^P)=0^+(1^{++})$ is produced from the $D\bar{D}^*$
interaction, which corresponds to the isoscalar particle $X(3872)$. No
bound state  related to the  $Z_c(3900)$
 was found. In lattice calculations, a candidate
X(3872) state was observed while  the possibility of a
shallow bound state related to the $Z_c(3900)$ was not supported~\cite{Chen:2014afa,Lee:2014uta,Prelovsek:2013xba}.


Generally, the theoretical studies suggest that there exists a
bound state relevant to the $X(3872)$  from the $D\bar{D}^*$ interaction, while the existence of a bound state relevant to the $Z_c(3900)$ is disfavored.
Besides the tetraquark interpretation
\cite{Braaten:2013boa,Deng:2014gqa, Qiao:2013raa}, many authors proposed  an alternative
explanation that the structure $Z_c(3900)$ is simply
a kinematical effect \cite{Swanson:2014tra,Chen:2013coa}. In their opinion, such structure is not related to an S-matrix pole  and
therefore should not be interpreted as a state. In Ref.
~\cite{Guo:2014iya}, it was suggested that  in the elastic channel the kinematic threshold cusp cannot produce a narrow
peak in the invariant mass distribution in
contrast with a genuine S-matrix pole, which can be used to distinguish kinematic cusp effects from genuine
 poles.


In scattering theory, a peak structure in the experiment can be
related to not only a bound state below threshold but also a
resonance above threshold. Both the bound state and resonance are from an attractive interaction, but the former needs stronger attraction \cite{Taylor1972}. For example, a popular interpretation of the $\Lambda(1405)$ is
a dynamically generated state with a two-pole structure \cite{Oller:2000fj,Oset:1997it,
Jido2003}. It is interesting to note that the higher-energy $\bar{K} N$ channel has a stronger attraction to support a bound state, while the lower energy $\pi\Sigma$ channel shows a relatively weaker attraction, which is nevertheless strong enough to generate a resonance \cite{Hyodo:2011ur}. In all measured channels, the experimental mass of the $Z_c(3900)$ is higher than the threshold \cite{Agashe:2014kda}, which also indicates that it may be a resonance instead of a bound state. Hence, it is
interesting to study the possibility of interpreting the $Z_c(3900)$ as a resonance above the $D\bar{D}^*$ threshold instead of a bound state.

In the literature, study about the possibilities of the $XYZ$ particles as a resonance has been scarce. There exist many works to study the bound state from meson-meson interactions by the potential model, the QCD sum rule and the lattice calculation. However, it is difficult to deal with a resonance with available QCD sum rule or lattice technology. In Ref. \cite{Aceti:2014uea},  the poles of a $T$ matrix were searched in
the local hidden gauge approach with heavy quark spin symmetry, but only the bound state from the $D\bar{D}^*$
interaction was studied. In this work, the method in Ref.~\cite{He:2014nya} will be
extended to search the poles for both bound states and resonance, and the heavy quark effective theory will be used to describe the $D\bar{D}^*$ interaction with light meson exchanges plus short-range $J/\psi$ exchange.

This work is organized as follows: In the next section a
theoretical frame is developed based on a quasipotential approximation
of the Bethe-Salpeter equation.
In Sec. III, the potential kernel with light meson exchange and $J/\psi$ exchange is derived with the
help of the effective Lagrangian from the heavy quark effective
theory. The numerical results are given in Sec IV. In the last
section, a summary is given.

\section{Scattering amplitude}

\begin{figure}[h!]
\includegraphics[width=0.48\textwidth]{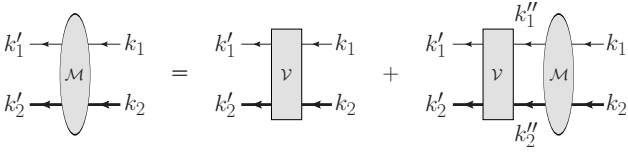}
\caption{The diagram for the Bethe-Salpeter equaiton. The thick and thin lines are for particles 1 and 2, respectively. \label{Fig: BS}}
\end{figure}

The scattering amplitude can be obtained through solving
the Bethe-Salpeter equation. The general form of the Bethe-Salpeter equation for the scattering amplitude as
shown in Fig.~\ref{Fig: BS} reads,
\begin{eqnarray}
&&{\cal M}(k'_1k'_2,k_1k_2;P)\nonumber\\&=&{\cal
V}(k'_1k'_2,k_1k_2;P)+\int\frac{d^4
k''}{(2\pi)^4}\nonumber\\
&\cdot&
{\cal
V}(k'_1k'_2,k''_1k''_2;P)G(k''_1k''_2){\cal
M}(k''_1k''_2,k_1k_2;P), \label{Eq: BS}
\end{eqnarray}
where ${\cal V}$ is the potential kernel and $G$ is the product of the propagators
for two constituent particles. Here the momentum of the system $P=k_1+k_2=k'_1+k'_2$,  and the relative momentum $k''=(k''_2-k''_1)/2$.

In this work, a quasipotential approximation will be applied to reduce
the four-dimensional Bethe-Salpeter equation to a three-dimensional one. Here the
covariant spectator theory will be applied as shown in Appendix A, in
which the heavier meson, particle 2, is put on shell. After multiplying
the polarized vector on both sides of the equation, we have
\begin{eqnarray}
{\cal M}_{\lambda',\lambda}({\bm p}',{\bm p})&=&{\cal
V}_{\lambda'\lambda}({\bm p}',{\bm p})+\sum_{\lambda''}\int\frac{d^3
p''}{(2\pi)^3}\nonumber\\
&\cdot&
{\cal
V}_{\lambda'\lambda''}({\bm p}',{\bm p}'')G_0({\bm p}''){\cal
M}_{\lambda''\lambda}({\bm p}'',{\bm p}),\quad \label{Eq: BS}
\end{eqnarray}
where ${\bm p}$, ${\bm p}'$ and ${\bm p}''$ are the momenta of constituent 2.
And the potential ${\cal V}_{\lambda'\lambda}({\bm p}',{\bm p})=\epsilon^{\mu*}_{\lambda'}
({\bm p}'){\cal V}_{\mu\nu}({\bm p}',{\bm p})\epsilon^\nu_\lambda({\bm p})$.

To reduce the Bethe-Salpeter equation to a one-dimensional equation, we apply the
partial wave expansion as shown in Appendix B.
The partial wave Bethe-Salpeter equation  with fixed
parity reads as,
\begin{eqnarray}
{\cal M}^{J^P}_{\lambda'\lambda}({\rm p}',{\rm p})
&=&{\cal V}^{J^P}_{\lambda',\lambda}({\rm p}',{\rm
p})+\sum_{\lambda''}\int\frac{{\rm
p}''^2d{\rm p}''}{(2\pi)^3}\nonumber\\
&\cdot&
{\cal V}^{J^P}_{\lambda'\lambda''}({\rm p}',{\rm p}'')
G_0({\rm p}''){\cal M}^{J^P}_{\lambda''\lambda}({\rm p}'',{\rm
p}).\quad\quad \label{Eq: BS_PWA}
\end{eqnarray}
Note that the sum extends only over non-negative $\lambda''$, and a factor
$1/\sqrt{2}$ has been included in the scattering amplitude and potential
for zero helicity.
The potential is defined as
\begin{eqnarray}
{\cal V}_{\lambda'\lambda}^{J^P}({\rm p}',{\rm p})
&=&2\pi\int d\cos\theta
~[d^{J}_{\lambda\lambda'}(\theta)
{\cal V}_{\lambda'\lambda}({\bm p}',{\bm p})\nonumber\\
&+&\eta d^{J}_{-\lambda\lambda'}(\theta)
{\cal V}_{\lambda'-\lambda}({\bm p}',{\bm p})],
\end{eqnarray}
where the momenta are chosen as $k_1=(W-E,0,0,-{\rm p})$,
$k_2=(E,0,0,{\rm p})$  and $k'_1=(W-E',-{\rm p}'\sin\theta,0,-{\rm p}'\cos\theta)$,
$k'_2=(E',{\rm p}'\sin\theta,0,{\rm p}'\cos\theta)$ with ${\rm p}=|{\bm p}|$ in order to avoid confusion with the
four-momentum $p$.

The above equation can be related to the Lippmann-Schwinger equations used by Oset $et\ al.$, if the potential kernel
${\cal  V}$ is only dependent on  the square of the momentum of the system $s=P^2$ and the $G_0$ is chosen as the one
used in Ref. \cite{Oset:1997it}. A cutoff regularization has been introduced in the  integration of the propagator in  Ref. \cite{Oset:1997it}, and it is related to a dimensional regularization. Since in our formalism, the integration is on the potential also the only cutoff  regularization is practical here. In this work we will adopt an  exponential
regularization instead of cutoff
regularization  by introducing a form factor in the propagator as
\begin{eqnarray}
	G_0({\rm p})\to G_0({\rm p})\left[e^{-(k_1^2-m_1^2)^2/\Lambda^4}\right]^2.\label{Eq: FFG}
\end{eqnarray}
Here particle 2 is not involved in the form factor due to its
on-shell-ness. The exponential
regularization used here can be seen as a softer version of the cutoff regularization  in Ref. ~\cite{Oset:1997it}, where the momentum $\rm p$ is cutoff at a certain value ${\rm p}_{max}$. The cutoff $\Lambda$ plays an analogous role to the cutoff ${\rm p}_{max}$ of cutoff
regularization. It can also be understood as a form factor in exponential form for the charmed mesons to reflect the internal structure of the hadron and to make the integration convergent. It is consistent with the OBE model where a form factor is usually added for the off-shell particle.

\section{The potential}

In Ref.~\cite{He:2014nya}, it was explained explicitly how to construct a potential for states with definite
isospin under SU(3) symmetry with the
corresponding flavor wave functions~\cite{Sun:2011uh}
\begin{eqnarray}
|Z_{D\bar{D}^*}^+\rangle_{I=1}&=&\frac{1}{\sqrt{2}}\big(|D^{*+}\bar{D}^0\rangle+c|D^+\bar{D}^{*0}\rangle\big),\nonumber\\
|Z_{D\bar{D}^*}^-\rangle_{I=1}&=&\frac{1}{\sqrt{2}}\big(|D^{*-}\bar{D}^0\rangle+c|D^-\bar{D}^{*0}\rangle\big),\nonumber\\
|Z_{D\bar{D}^*}^0\rangle_{I=1}&=&\frac{1}{2}\Big[\big(|D^{*+}D^-\rangle-|D^{*0}\bar{D}^0\rangle\big)\nonumber\\
&+&c\big(|D^+D^{*-}\rangle-|D^0\bar{D}^{*0}\rangle\big)\Big],\label{Eq: wf1}\nonumber\\
|Z_{D\bar{D}^*}^0\rangle_{I=0}&=&\frac{1}{2}\Big[\big(|D^{*+}D^-\rangle+|D^{*0}\bar{D}^0\rangle\big)\nonumber\\
&+&c\big(|D^+D^{*-}\rangle+|D^0\bar{D}^{*0}\rangle\big)\Big],  \label{Eq: wf2}
\end{eqnarray}
\normalsize where $c=\pm$ corresponds to $C$-parity $C=\mp$ respectively. For the isovector state the $c$ is related to the $G$-parity.

Basically, the strong interaction should be described by gluon and quark freedoms. If the distance between two hadrons is large, the interaction will appear as meson exchange, that is, the OBE model. Many efforts have been made to study the connection between  QCD and the OBE model. For example, in Refs. \cite{Simonov:2011cm,Danilkin:2011sh} after integrating out quark degrees of
freedom in the effective Lagrangian, one obtains the chiral effective Lagrangian
for mesons. Though there is still not particularly convincing about the connection, as a phenomenological model the people's confidence about the OBE model arises from the successes of its applications to the deuteron,  such as the CD-Bonn model \cite{Machleidt:1987hj} and the Gross model \cite{Gross:2010qm}, and the constituent quark model by Riska and Glozman \cite{Glozman:1995fu}.  If the X(3872) is interpreted as a molecular state, it should have a radius of about 7~fm, as estimated by Close and Page~\cite{Close:2003sg}. For a resonance above the threshold, it is reasonable to assume the interaction is at a large distance. Hence, the OBE model will be adopted to describe the $D\bar{D}^*$ interaction in this work.

It seems strange to include the vector-meson exchanges, which mainly take effect at short distances.  However, we should remind the reader that a short-distance interaction does not mean an interaction only at short distances. The vector-meson exchange will have some remnant at long distances. It is also can be understood as indicating that there is some possibility that the two hadrons are close to each other. If the interaction from the vector-meson exchange is large enough, it will have a considerable remnant at long distances. Based on this consideration, the vector-meson exchanges are also included in the OBE model.  At short distances, the studies in the constituent quark model suggests the gluon exchange contribution can be replaced by the contribution from vector-meson exchanges \cite{Dai:2003dz,He:2003vi}. Therefore, vector-meson exchanges will be included in the calculation in this work  instead of gluon exchange.

\begin{figure}[h!]
\includegraphics[width=0.48\textwidth]{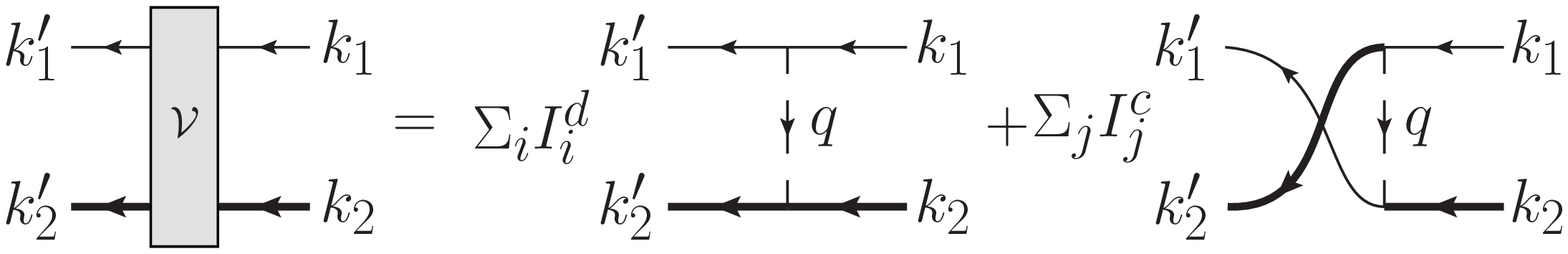}
\caption{The potential including both direct and cross diagrams. $I^d_i$ and $I^c_i$ are the isospin factors for direct and cross diagrams, respectively. }
\label{Fig: potential}
\end{figure}

There exist two types of diagram namely, the direct diagram and the cross diagram (see Fig.~\ref{Fig: potential}), as in a conventional OBE potential model \cite{Sun:2011uh,Sun:2012zzd}. In the cross diagram, the final particles are alternated. With such alternation, the propagator is the same for different components in a SU(3) state, such as $|D^{*+}\bar{D}^0\rangle$ and  $|D^+\bar{D}^{*0}\rangle$ for the positive-charge state, so that the equations for different components are reduced to one equation under S(3) symmetry~\cite{He:2014nya}.

To write the potential, we adopt the effective Lagrangians of
the pseudoscalar and vector mesons with heavy flavor mesons from the heavy quark effective theory ~\cite{Colangelo:2003sa,Casalbuoni:1996pg}.
The potential kernels ${\cal V}$ from vector-meson ($\mathbb{V}$) exchange, pseudoscalar-meson ($\mathbb{P}$) exchange, and  scalar-meson ($\sigma$) exchange have been given in Ref.~\cite{He:2014nya} as
\begin{eqnarray}
{\cal V}^{Direct}_{\mathbb{V}}&=&i\frac{\beta^2g^2_V}{2}
	\frac{(k_2+k'_2)\cdot(k_1+k'_1)~\epsilon_2\cdot\epsilon'_2}{q^2-m_\mathbb{V}^2},\nonumber\\
	{\cal V}^{Direct}_{\sigma}&=&i4g^2_\sigma
m_{P}m_{P^*}\frac{\epsilon_2\cdot\epsilon'_2}{q^2-m^2_\sigma},\nonumber\\
{\cal V}^{Cross}_\mathbb{V}&=&i2\lambda^2
g^2_V\varepsilon_{\lambda\alpha\beta\mu}(k_2+k'_1)^\lambda
q^\alpha
\epsilon_2^\mu~\nonumber\\
&\cdot&\varepsilon_{\lambda'\alpha'\beta\nu}(k_1+k'_2)^{\lambda'}q^{\alpha'}\epsilon'^\nu_2
\frac{1}{q^2-m_\mathbb{V}^2},\nonumber\\
	{\cal
	V}^{Cross}_{\mathbb{P}}&=&i\frac{4g^2m_Pm_{P^*}}{f^2_\pi}
	\frac{k'_1\cdot\epsilon_2 k_1\cdot\epsilon'_2}{q^2-m^2_\mathbb{P}},
\end{eqnarray}
where the momenta $k^{(')}_{1,2}$ and $q$ are defined as in Fig.~\ref{Fig: potential}. Here The $\epsilon^{(')}_2$ is the polarization vector for the  initial or final particle 2. $m_\mathbb{P}$, $m_\mathbb{V}$ and $m_\sigma$ are the masses for the exchange pseudoscalar, vector, and sigma mesons. In the OBE model, the masses of all mesons used are from the PDG~\cite{Agashe:2014kda}. The $\sigma$ meson is seen as a particle with a mass 475 MeV, which is the center value suggested by the PDG.

In the OBE model, we will adopt the physical values of the coupling constants. How  to determine the coupling constants has been discussed in the heavy quark effective theory \cite{Casalbuoni:1996pg}. The coupling constant $g$ for the pseudoscalar exchange is extracted from the experimental width
of $D^{*+}$ as $g=0.59$~\cite{Isola:2003fh}. The parameter $\beta$ for the vector
meson was fixed as $\beta = 0.9$ by the vector-meson
dominance mechanism and $\lambda=0.56$ GeV$^{-1}$ was obtained
by comparing the form factor obtained by lattice
QCD with the one  calculated by the
light-cone sum rule ~\cite{Colangelo:2003sa}. The coupling constant $g_\sigma = g_\pi/(2\sqrt{6})$ with $g_\pi = 3.73$ was given in Ref.~\cite{Falk:1992cx}.

A form factor is introduced to compensate the
off-shell effect of the exchange meson  as  $f(q^2)=(\frac{\Lambda^2}{\Lambda^2-q^2})^4$. It is different from the one used in the propagator [see Eq. (\ref{Eq: FFG})]
, which is usually used in a case where the off-shell particle has possibility $q^2=\Lambda^2$, to avoid an unnecessary pole arising from the form factor. The cutoff can be related to the radius of the hadron as
\begin{eqnarray}
 r^2=-\frac{6}{F(0)}\frac{d}{d q^2} F(q^2)|_{q^2=0}.\label{Eq: radius}
\end{eqnarray}
With Eq.~(\ref{Eq: radius}), the cutoffs in two types of form factors have relation $\Lambda\approx5/r$.  If we assume the mesons have radii of about 0.5~fm, the cutoff is about 2 GeV.
Here the momenta for the exchange mesons are defined as $q=k'_2-k_2$ and $q=k'_1-k_2$ for direct and cross diagrams, respectively.
In the propagator of the meson exchange we make a replacement $q^2\to-|q^2|$ to remove the unphysical singularities, as in Ref.~\cite{Gross:2008ps}.

In Ref. \cite{Aceti:2014uea}, it was suggested that the $J/\psi$ exchange is important in the $D\bar{D}^*$ interaction. And the potential was written  in
the local hidden gauge approach with heavy quark spin symmetry. In this work,  the couplings of heavy-light charmed mesons to $J/\psi$ are written with help of the heavy quark effective theory as \cite{Casalbuoni:1996pg,Oh:2000qr}
\begin{eqnarray}
	{\cal L}_{D^*_{(s)}\bar{D}^*_{(s)}J/\psi}&=&-ig_{D^*_{(s)}D^*_{(s)}\psi}\big[\psi \cdot \bar{D}^*\overleftrightarrow{\partial}\cdot D^*\nonumber\\
&-&
\psi^\mu \bar D^* \cdot\overleftrightarrow{\partial}^\mu {D}^* +
\psi^\mu \bar{D}^*\cdot\overleftrightarrow{\partial} D^{*\mu} ) \big], \nonumber \\
{\cal L}_{D_{(s)}^*\bar{D}_{(s)}J/\psi}&=&
g_{D^*_{(s)}D_{(s)}\psi} \,  \, \epsilon_{\beta \mu \alpha \tau}
\partial^\beta \psi^\mu (\bar{D}
\overleftrightarrow{\partial}^\tau D^{* \alpha}+\bar{D}^{* \alpha}
\overleftrightarrow{\partial}^\tau D) \label{matrix3}, \nonumber \\
{\cal L}_{D_{(s)} \bar{D}_{(s)}J/\psi} &=&
ig_{D_{(s)}D_{(s)}\psi} \psi \cdot
\bar{D}\overleftrightarrow{\partial}D,
\end{eqnarray}
where the couplings
are related to a single parameter $g_2$ as
\begin{eqnarray}
\frac{g_{D^*D^*\psi} }{m_{D^*}}= \frac{g_{D_{(s)}D_{(s)}\psi}}{m_D}= g_{D^*_{(s)}D_{(s)}\psi}= 2 g_2 \sqrt{m_\psi },
\end{eqnarray}
with $g_2={\sqrt{m_\psi}}/({2m_Df_\psi})$ with $f_\psi=405$ MeV.

With the above Lagrangians, the potential kernel for $J/\psi$ exchange is written as,
\begin{eqnarray}
{\cal
	V}^{Direct}_{J/\psi}&=&-ig_{D^*D^*J/\psi}g_{DDJ/\psi}[\epsilon'_2\cdot(k_1+k'_1)~\epsilon_2\cdot(k_2+k'_2)\nonumber\\
	&+&\epsilon'_2\cdot(k_2+k'_2)~\epsilon_2\cdot(k_1+k'_1)\nonumber\\
	&-&(k_2+k'_2)\cdot(k_1+k'_1)~\epsilon'_2\cdot\epsilon_2]\frac{1}{q^2-m_{J/\psi}^2},\nonumber\\
{\cal
	V}^{Cross}_{J/\psi}&=&ig_{DD^*J/\psi}^2
~\varepsilon_{\lambda\alpha\beta\mu}(k_2+k'_1)^\lambda
q^\alpha
\epsilon_2^\mu~\nonumber\\
&\cdot&\varepsilon_{\lambda'\alpha'\beta\nu}(k_1+k'_2)^{\lambda'}q^{\alpha'}\epsilon'^\nu_2\frac{1}{q^2-m_{J/\psi}^2}.
\end{eqnarray}

We would like to point out that the potentials obtained by the heavy quark effective theory are comparable to the ones obtained from the chiral Lagrangian in Ref.~\cite{Aceti:2014uea} after a nonrelativization. For example, the potential kernel for the $J/\psi$ exchange can be rewritten as
\begin{eqnarray}
{\cal
	V}^{Direct}_{J/\psi}&=&-\frac{g_{D^*D^*J/\psi}g_{DDJ/\psi}}{m_{J/\psi}^2}(k_2+k'_2)\cdot(k_1+k'_1)~{\bm \epsilon}'_2\cdot{\bm \epsilon}_2,\ \ \
\end{eqnarray}
with  ${g_{D^*D^*J/\psi}g_{DDJ/\psi}}/{m_{J/\psi}^2}=7.04$ in this work and \begin{eqnarray}
{\cal
	V}^{Direct}_{J/\psi}&=&-\frac{{\cal C}_{ij}}{4f^2}(k_2+k'_2)\cdot(k_1+k'_1)~{\bm \epsilon}'_2\cdot{\bm \epsilon}_2,\ \
\end{eqnarray}
with ${{\cal C}_{ij}}/{4f^2}=6.89$ in Ref.~\cite{Aceti:2014uea}.  The main difference between our work and Ref.~\cite{Aceti:2014uea} is that a form factor is added not only to the light pseudoscalar-meson exchanges  but also to the $J/\psi$ exchange, which will suppress the contribution from $J/\psi$ exchange.

The flavor factors $I_i^d$ and $I_i^c$ for direct and cross diagrams are presented in  Table \ref{flavor factor}. The cancellation of the $\rho$ and $\omega$ meson exchanges happens in the isovector sector as suggested by the isospin factor listed in Table~\ref{flavor factor}, which leads to a shortage of the short-range interaction in that sector if the $J/\psi$ exchange is absent.
\renewcommand\tabcolsep{0.21cm}
\renewcommand{\arraystretch}{1.7}

\begin{table}[hbtp!]
\begin{center}
\caption{The isospin factors $I_i^d$ and $I_i^c$ for direct and cross diagrams and different exchange mesons.
\label{flavor factor}}
\begin{tabular}{c|cccc|ccccc}\bottomrule[2pt]
&\multicolumn{4}{c|}{Direct diagram} &\multicolumn{5}{c}{Crossed diagram}\\\hline
&$\rho$ &$\omega$&$J/\psi$ &$\sigma$&$\pi$&$\eta$  &$\rho$ &$\omega$ &$J/\psi$ \\\hline
$[PP^*]^T$
&$-\frac{1}{2}$&$\frac{1}{2}$&1 &1 &$-\half c$
&$\frac{1}{6} c$
&$-\half c$  &$\half c$ &$c$\\
$[PP^*]^S$
&$\thalf$&$\half$& 1&1&$\frac{3}{2}c$ &
$\frac{1}{6}c$ & $\thalf c$
&$\frac{1}{2}c$ &$c$\\
\toprule[2pt]
\end{tabular}
\end{center}
\end{table}

\section{Numerical results}

In this work, we will search  the poles of the scattering amplitude ${\cal M}$ from the $D\bar{D}^*$ interaction which is described by the potential kernel obtained in the above section. The scattering amplitude is obtained by solving the Bethe-Salpeter equation.

\subsection{Numerical solution of the Bethe-Salpeter equation}

Before solving the one-dimensional partial wave Bethe-Salpeter equation numerically,
we need to deal with the pole in $G_0({\rm p})$ as (here the notation $J^P$ is omitted)
\begin{eqnarray}
	&&i{\cal M}({\rm p},{\rm p}')=i{\cal V}({\rm p},{\rm p}')\nonumber\\
	&+&\int\frac{{\rm p}''^2d {\rm p}''}{(2\pi)^3}
	i{\cal V}({\rm p},{\rm p}'') G_0({\rm p}'')i{\cal M}({\rm
	p}'',{\rm p'})\nonumber\\
	&-&i{\cal V}({\rm p},{{\rm p}}_o'')[\int\frac{d {\rm p}''}{(2\pi)^3}\frac{A({\rm p}''_o)}{
		{\rm p}''^2-{\rm p}''^2_o}+i\frac{{\rm p}''^2_o\delta\bar{G}_0({\rm p}''_o)}{8\pi^2}]\nonumber\\
&\cdot&i{\cal
M}({\rm p}''_o,{\rm p}')\theta(s-m_1-m_2),
\end{eqnarray}
with
\begin{eqnarray}
	&&A({\rm p}''_o)=[{\rm p}''^2({\rm p}''^2-{\rm p}''^2_o)G_0({\rm
	p}'')]_{ {\rm p}''\to{\rm p}''_o}=
-\frac{{\rm p}''^2_o}{2W},\nonumber\\
&&\delta \bar{G}_0({\rm p}'')\delta({\rm p}''-{\rm p}''_o)=\delta(G_0({\rm
p})'')=\frac{1}{4W{\rm p}''_o}\delta({\rm p}''-{\rm p}''_o),\quad\quad
\end{eqnarray}
with ${\rm p}''_o=\frac{1}{2W}\sqrt{[W^2-(M+m)^2][W^2-(M-m)^2]}$.

To solve the integral equation, we discrete the momenta ${\rm p}$,
${\rm p}'$, and ${\rm p}''$ by the Gauss quadrature with wight $w({\rm
p}_i)$ and have
\begin{eqnarray}
{M}_{ik}
&=&{V}_{ik}+\sum_{j=0}^N{ V}_{ij}G_j{M}_{jk},
\end{eqnarray}
where  $i$ is absorbed in $M$ or $V$. The discreted propagator is of a form
\begin{eqnarray}
	G_{j>0}&=&\frac{w({\rm p}''_j){\rm p}''^2_j}{(2\pi)^3}G_0({\rm
	p}''_j), \nonumber\\
G_{j=0}&=&-\frac{i{\rm p}''_o}{32\pi^2 W}+\sum_j
\left[\frac{w({\rm p}_j)}{(2\pi)^3}\frac{ {\rm p}''^2_o}
{2W{({\rm p}''^2_j-{\rm p}''^2_o)}}\right].\quad\quad
\end{eqnarray}

In this work, we will search the poles from the amplitude of  elastic scattering where the initial and final particles are on shell.  The scattering amplitude is
\begin{eqnarray}
	M=M_{00}=\sum_j[(1-{ V} G)^{-1}]_{0j}V_{j0}.\label{Eq: Scattering ampltiudes}
\end{eqnarray}

The pole can be searched by variation of $z$ to satisfy
\begin{eqnarray}
	|1-V(z)G(z)|=0,
\end{eqnarray}
where  $z=E_R+i\Gamma/2$ equals the meson-baryon energy $W$ at the real axis. Since
$z=\sqrt{m_1^2+{\rm p}^2}+\sqrt{m_2^2+{\rm p}^2}$, the $\rm p$-plane
corresponds to two Reimann sheets for $z$. A bound state is located
in the first Reimann sheet while a resonance is located in the second
Reimann sheet with Im(p)$<$0. Since only one channel is considered in this work, the bound state is located at the real axis  of the $z$ complex plane, while the resonance will deflect the real axis to the complex plane.

\subsection{Bound state from $D\bar{D}^*$ interaction with a $\Lambda$ scan}

There exist two types of pole, namely bound state and resonance. First, we will make a $\Lambda$ scan from 0.8~GeV to 4~GeV to find  the bound state from the $D\bar{D}^*$ interaction. The pole of a bound state from $D\bar{D}^*$ interaction is located at the real axis.

\renewcommand\tabcolsep{0.112cm}
\renewcommand{\arraystretch}{1.7}
\begin{table}[hbtp!]
\begin{center}
\caption{The position of the bound state from the $D\bar{D}^*$ interaction at the real axis  $Re(z)=W$ with a $\Lambda$ scan. The second and third  columns are for the full model. The results without the $J/\psi$ exchange are listed in the fourth and fifth columns and compared with the results in Ref. \cite{He:2014nya}. The $J/\psi$ (I) in the eighth and ninth columns and $J/\psi$ (II) in tenth and eleventh columns  are for the results from only $J/\psi$ exchange with and without form factor. The cutoff $\Lambda$ and  energy $W$ are in units of GeV. \label{Tab: bound state}
\label{diagrams}}
	\begin{tabular}{c|cccccccccc}\bottomrule[2pt]
& \multicolumn{2}{c}{Full model}& \multicolumn{2}{c}{No $J/\psi$}& \multicolumn{2}{c}{Ref. \cite{He:2014nya}}& \multicolumn{2}{c}{$J/\psi$ (I)}& \multicolumn{2}{c}{$J/\psi$ (II)}	 \\\hline
$I^G(J^{PC})$   &  $\Lambda$ & $W$ &  $\Lambda$ & $W$ &  $\Lambda$ & $W$ &  $\Lambda$ & $W$ &  $\Lambda$ & $W$\\\hline
$0^-(0^{--})$& --  & -- & -- & -- & -- & -- & -- & --   &--&--\\
$0^+(0^{-+})$& --  & -- & -- & -- & -- & -- & -- & --  &--&-- \\
$0^-(1^{--})$& --  & -- & -- & -- & -- & -- & -- & --  & -- & --\\
$0^+(1^{-+})$& --  &--  & -- & -- & -- & -- & -- & --  & -- & --\\
$0^-(1^{+-})$& 1.0  & 3.864 & 1.0 & 3.868 & 1.3 & 3.876 &--&-- &2.5 & 3.867\\
             & 1.2  & 3.848 & 1.2 & 3.854 & 1.4 & 3.870 &--&-- &2.6 & 3.850\\
$0^+(1^{++})$& 1.9  & 3.873 & 1.9  &3.875 & 2.0 & 3.876 &-- &-- & 3.9 & 3.875\\
             & 2.4  & 3.871 & 2.4  &3.874 & 2.4 & 3.872 &-- &-- & 4.0 & 3.836\\\hline
$1^+(0^{-})$& --  &--  & --& --& --& -- &--&--&--&--\\
$1^-(0^{-})$& --  &--  & --& --&-- &-- &--&--&--&--\\
$1^+(1^{-})$& --  &--      & -- & -- & -- & --&--&--&--&--\\
$1^-(1^{-})$& --  &--      & -- & -- & -- & -- &--&--&--&--\\
$1^+(1^{+})$& 3.0 & 3.874  & -- & -- & -- & -- &--&--&2.4 &3.875\\
            & 3.3 & 3.858  & -- & -- & -- & -- &--&--&2.5 &3.867\\
$1^-(1^{+})$& --  & --     & -- & -- & -- & -- &--&--&--&--\\
\toprule[2pt]
\end{tabular}
\end{center}

\end{table}

One can find that the $J/\psi$ exchange plays a more important role in the isoscalar sector than in the isovector sector, where the short-range interaction is absent due to the cancellation between $\rho$ and $\omega$ exchanges in the isoscalar sector, as shown in Table~\ref{flavor factor}.  As expected, the results without the $J/\psi$ exchange are also close to those obtained from  the solution of a Bethe-Salpeter equation for the vertex in Ref. \cite{He:2014nya}. The results with only the $J/\psi$ exchange are also listed in Table~\ref{Tab: bound state}. There is no bound state found for the $J/\psi$ exchange with form factor [labeled as $J/\psi$ (I)], while if the form factor is removed  [labeled as $J/\psi$ (II)], the bound state is found as in the full model, which indicates the form factor weakens the contribution from the $J/\psi$ exchange.

In the isoscalar vector, there exist  hidden
charmed bound states with $I^G(J^{PC})=0^-(1^{+-})$ and $0^-(1^{++})$ with  cutoffs about 1~GeV and 2~GeV, respectively.
The $D\bar{D}^*$ bound state with $0^+(1^{++})$ can be
related to the $X(3872)$. In the isovector sector, a bound  state with $1^+(1^+)$  can be found with a larger cutoff at about 3 GeV.

\subsection{The $Z_c(3900)$ as a resonance}

\begin{figure}[h!]
\includegraphics[bb=110 30 340 300,clip, scale=1.07]{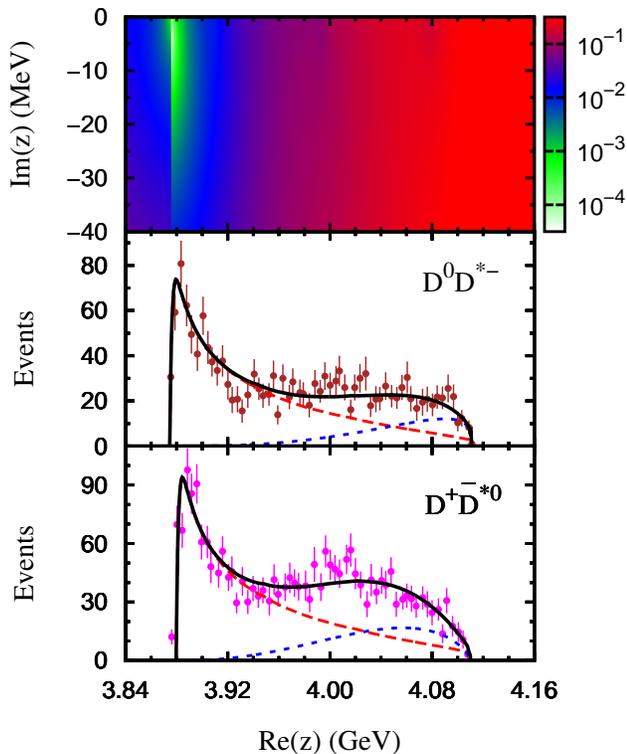}
\caption{(color online) The $|1-G(z)V(z)|$ for $I^G(J^{PC})=1^+(1^+)$ in the complex energy plane (upper panel) and the $D^0D^{*-}$ (middle panel) and $D^+ \bar{D}^{*0}$ (lower panel) invariant mass spectra. The abscissa axis $Re(z)$ represents the
corresponding $D\bar{D}^*$ invariant mass $W$ in units of GeV.
The solid, dotted, and dashed lines are for results with
the full model, the resonance contribution,  and
the background, respectively. The
experimental results are from BESIII~\cite{Ablikim:2013xfr}.The theoretical results are normalized to the experiment.  }
\label{Fig: Res}
\end{figure}

In physics, the cutoff in the $D\bar{D}^*$ interaction should be same for different quantum numbers. Usually, a decrease of the cutoff will lead to a weaker  interaction, and vice versa.
As stated in the scattering theory, when interaction weakens, the bound state runs to the threshold and becomes a resonance if the interaction is still strong enough.  If the cutoff is increased to 3~GeV, the pole with $0^+(1^{++})$ will move to about 3650~MeV, which is very far from the experimental mass of $X(3872)$. Hence, we decrease the cutoff $\Lambda$ for $1^+(1^+)$ from
 about 3 GeV to 2.4~GeV with which a bound state which has quantum number $0^{+}(1^{++})$ is found at 3.871 GeV.  A pole with $1^+(1^+)$ is produced slightly higher than the $D\bar{D}^*$ threshold with a cutoff $\Lambda=2.4$ GeV as shown in the upper panel of Fig.~\ref{Fig: Res}. It is located at $z=3876+i5$ MeV  and can be identified with the charged charmonium-like state $Z_c(3900)$ observed in BESIII.

Obviously, the position of the pole is below the experiment masses of $Z_c(3900)$, 3883.9$\pm1.5\pm4.2$~MeV in the $D\bar{D}^*$ channel and $3899\pm3.6\pm4.9$~MeV in the $\pi^-J/\psi$ channel \cite{Agashe:2014kda}. The experimental mass is obtained by fitting the invariant mass spectrum. As shown in the middle and lower panels of Fig.~\ref{Fig: Res}, the invariant mass spectrum of the $D\bar{D}^*$ channel is presented and
compared with the experimental results released by the BESIII Collaboration  \cite{Ablikim:2013xfr}.

The invariant mass
distribution is given approximately as \cite{Hyodo:2003jw,Aceti:2014uea}
\begin{eqnarray}
	\frac{d\sigma}{dW}=C|{M}^{1^+}|^2\lambda^\half(W^2,M^2,m^2)\lambda^\half
	(\tilde{W}^2,W^2,m_3^2)/W, \label{Eq: mass spectrtum}
\end{eqnarray}
where $M^{1^+}$ is the scattering amplitude obtained from the Bethe-Salpeter eqaution as defined in Eq. (\ref{Eq: Scattering ampltiudes}), $\tilde{W}$ is total energy of the process, and $m_3$ is the
third final particle $\pi$ \cite{Hemingway:1984pz,Aceti:2014uea}. A background spectrum  is also included as in Refs.~\cite{Ablikim:2013xfr,Aceti:2014uea},
\begin{eqnarray}
f_{bkg}(W)= a(W-M_{min})^c (M_{max}-W)^d, \label{Eq: background}
\end{eqnarray}
where $M_{min}$ and $M_{max}$ are the minimum and maximum kinematically allowed masses. The general constant $C$ in Eq.~(\ref{Eq: mass spectrtum}), and the general constant  $a$ and exponents $c$ and $d$ in Eq.~(\ref{Eq: background})
are free parameters adjusted to reproduce the experimental data.

The experimental data are well reproduced with the resonance contribution plus a background. A peak is found at 3881~MeV which is higher than the position of the pole of the resonance but closer to the experimental results 3883.9$\pm1.5\pm4.2$~MeV in the $D\bar{D}^*$ channel  \cite{Ablikim:2013xfr}. We would like to note that due to the pole's nearness to the threshold the contribution from the $Z_c(3900)$  in our model does not have a standard Breit-Wigner form, which was adopted  in the experimental fitting ~\cite{Ablikim:2013xfr}. The nearness also results in a relatively large tail of the resonance at higher energies. These differences lead to a background contribution, which is quite different from the one in Ref.~\cite{Ablikim:2013xfr} to be used to reproduce the invariant mass spectrum.

\section{Summary}

In this work, the $D\bar{D}^*$ interaction is studied within the OBE model, including the contribution from  light meson
 exchanges plus
the short-range $J/\psi$ exchange.  The scattering amplitude is calculated
within a Bethe-Salpeter equation approach, and the poles near the  $D\bar{D}^*$
threshold are searched. It is found that the charged charmonium-like state
$Z_c(3900)$ can be interpreted a resonance above the threshold from the $D\bar{D}^*$
interaction.

In the isoscalar sector, the poles
are found under the $D\bar{D}^*$ threshold and have the quantum
numbers $I^G(J^{PC})=0^-(1^{+-})$ and $0^+(1^{++})$. If the $J/\psi$
exchange is excluded, the position of the pole is almost unaffected.
In the isovector sector, where the short-range contributions from
$\omega$ and $\rho$ exchanges are canceled, a bound state is found  with $I^G(J^P)=1^+(1^+)$. It disappears if the $J/\psi$
exchange is removed. The results show that the short-range $J/\psi$
exchange, which is not included in the conventional one-boson exchange
model, is important to provide an attractive interaction to produce the pole in the isovector sector.

If a cutoff $\Lambda=2.4$ GeV is
adopted with which a pole carrying the quantum number of the $X(3872)$  is
produced at an energy of about 3871 MeV, the pole for the bound state with
$1^+(1^+)$ runs across the threshold to a second Rienman sheet and becomes a resonance at $3876+i5$ MeV, which can be identified with the $Z_c(3900)$.  The line shape of the invariant mass spectrum in
the $D\bar{D}^{*}$ channel is also investigated and the experimental
results by the BESIII Collaboration can be reproduced. A peak is found in the $D\bar{D}^*$ invariant mass spectrum at about 3881 MeV, which is higher than the resonance pole but closer to the experimental values.

\emph{Note}: Just after the paper was submitted, we noticed a related work released in arXiv,
in which the author also found a second-sheet pole in studying elastic $D \bar{D}^*$ scattering\cite{Zhou:2015jta}.

\acknowledgements
This project is partially supported by the Major State
Basic Research Development Program in China (No. 2014CB845405),
the National Natural Science
Foundation of China (Grants No. 11275235, No. 11035006)
and the Chinese Academy of Sciences (the Knowledge Innovation
Project under Grant No. KJCX2-EW-N01).

\appendix
\section{The quasipotential approximation of the Bethe-Salpeter equation}

It
is popular to reduce the Bethe-Salpeter equation from a four-dimensional integral equation to a three-dimensional equation by quasipotential
approximation. In principle, 
infinite choices can be applied to make the quasipotential approximation. The
popular methods used in literature include the BSLT
approximation, the K-matrix method,  the instantaneous
approximation, and the covariant spectator
theory (CST) ~\cite{Blankenbecler:1965gx,Shklyar:2004ba,Guo:2007mm, Gross:1991pm,Gross:2008ps,He:2014nya,He:2015yva,Nieuwenhuis:1996mc}.

The reduced propagator $G$ under quasipotential approximations should satisfy the unitary condition,  i.e., the relation
\begin{eqnarray}
G-G^\dag=2\pi i\delta((\eta_1(s)
P+k)^2-m^2)\delta((\eta_2(s) P-k)^2-m^2),\ \ \
\end{eqnarray}
where $k_{1}=\eta_1(s)P+k$, $k_2=\eta_2(s)-k$, and $m_{1,2}$ are the momenta and mass of constituents 1 and 2 with $\eta_1(s)+\eta_2(s)=1$ and $s=P^2$.
One can define
$\eta_{1,2}=\epsilon_{1,2}/(\epsilon_1+\epsilon_2)$ with $\epsilon_{1,2}(s)=(s+m_{1,2}^2-m_{2,1}^2)/2\sqrt{s}$.
Now we have many choices to write the propagator.
The most popular form of propagator is~\cite{Hung:2001pz,Zakout:1996md,He:2013oma}
\begin{eqnarray}
	G&=&2\pi\int \frac{ds'}{s'-s+i\epsilon} h(s',s)\nonumber\\&\cdot&
	\delta([\eta'_1(s')P'+k]^2-m_1^2)~\delta([\eta'_2(s')P'-k]^2-m_2^2),\ \ \label{Eq: G0int}
\end{eqnarray}
where $P'=\sqrt{s'/s}P$.
It is random to some extent to choose $h(s',s)$ . The $h$ function is chosen as  $h(s'-s)=1$ with $\eta'(s')=\eta(s')$ for the BSLT formalisms,.
In the CST, $h(s',s)=(\sqrt{s'}+\sqrt{s})/\sqrt{s'}$, with
$\eta'_1(s')=\eta_1(s)\sqrt{s/s'}$ and
$\eta'_2(s')=1-\eta_1(s)\sqrt{s/s'}$.

Eqation~(\ref{Eq: G0int}) is quite far from exhausting all possible three-dimensional reductions. For example, the widely used instantaneous approximation has
\begin{eqnarray}
	G&=&\int d k_0\frac{-1}{(k_1^2-m_1^2+i\epsilon)(k_2^2-m_2^2+i\epsilon)}\nonumber\\
	&=&i\pi\frac{(E_1({\bm p})+E_2({\bm p}))/E_1({\bm p})E_2({\bm p})}{W^2-(E_1({\bm p})+E_2({\bm p}))^2},
\end{eqnarray}
which satisfies the unitary condition also.

As in
Ref.~\cite{He:2014nya}, the covariant spectator
theory~\cite{Gross:1991pm,Gross:2008ps} is adopted to make the quasipotential approximation to
reduce the four-dimensional Bethe-Salpeter equation to a three-dimensional
equation in the current work.  Written down in
the center-of-mass frame where $P=(W,{\bm 0})$, the propagator in the CST is
\begin{eqnarray}
	G&=&2\pi i\frac{\delta^+(k_2^{~2}-m_2^{2})}{k_1^{~2}-m_1^{2}}
	\nonumber\\&=&2\pi
	i\frac{\delta^+(k^{0}_2-E_2({\bm p}))}{2E_2({\bm p})[(W-E_2({\bm
p}))^2-E_1^{2}({\bm p})]},
\end{eqnarray}
where $k_1=(k_1^{0},-\bm
p)=(W-E_2({\bm p}),-\bm p)$ and$k_2=(k_2^{0},\bm
p)=(E_2({\bm p}),\bm p)$ with $E_{1,2}({\bm p})=\sqrt{
m_{1,2}^{~2}+|\bm p|^2}$. In our formalism a definition $G_0=G/2\pi$ is used for convenience.

\section{The partial wave expansion and the amplitudes with fixed parity}

The partial wave expansion of the scattering amplitude ${\cal M}$ in Eq. (\ref{Eq: BS})
is \cite{Chung}
\begin{eqnarray}
{\cal M}_{\lambda'\lambda}({\bm p}',{\bm p})&=&
	\sum
	 _{J\lambda_R}{\frac{2J+1}{4\pi}}D^{J*}_{\lambda_R,\lambda'}(\phi',\theta',0){\cal
	V}^J_{\lambda'\lambda,\lambda_R}({\rm p}',{\rm p})\nonumber\\
	&\cdot&D^{J}_{\lambda_R,\lambda}(\phi,\theta,0),
\end{eqnarray}
where $J$ is the angular momentum for the partial wave  considered and $D^{J*}_{\lambda_R,\lambda}(\phi,\theta,0)$ is the rotation matrix with $\lambda_R$ being the helicity of the bound state. The potential in the partial wave expansion equation (\ref{Eq: BS_PWA}) is
\begin{eqnarray}
	{\cal V}_{\lambda'\lambda}({\bm p}',{\bm p})&=&
	\sum
	 _{J\lambda_R}{\frac{2J+1}{4\pi}}D^{J*}_{\lambda_R,\lambda'}(\phi',\theta',0){\cal
	V}^J_{\lambda'\lambda,\lambda_R}({\rm p}',{\rm p})\nonumber\\ &\cdot&
	D^{J}_{\lambda_R,\lambda}(\phi,\theta,0).
\end{eqnarray}
Without loss of the generality, we choose the scattering to be in the $xz$ plane, the potential is written as
\begin{eqnarray}
{\cal V}_{\lambda'\lambda}^J({\rm p},{\rm p}')=2\pi\int d\cos\theta d^{J}_{\lambda\lambda'}(\theta_{p',p})
{\cal V}_{\lambda'\lambda}({\bm p}',{\bm p}),
\end{eqnarray}
where the momenta are chosen as $k_1=(W-E,0,0,-{\rm p})$,
$k_2=(E,0,0,{\rm p})$  and $k'_1=(W-E',-{\rm p}'\sin\theta,0,-{\rm p}'\cos\theta)$,
$k'_2=(E',{\rm p}'\sin\theta,0,{\rm p}'\cos\theta)$ with ${\rm p}=|{\bm p}|$ in order to avoid confusion with the
four-momentum $p$. The particle helicities $\lambda$ are the projections of the spin $s$ on the direction of motion of the particle. Here and hereafter, the individual helicities are omitted where redundant and the states are only labeled
by the total helicities $\lambda$, $\lambda'$ and $\lambda''$.  Thus, once in the center-of-mass system the $z$-axis is chosen along the three-momentum of the incoming particle 1, and one has $\lambda_1 =s_1$ for final state particle 1 and $\lambda_2=-s_2$ for final state particle 2.

The amplitudes with definite parity can be constructed as ~\cite{G.Penner}
\begin{eqnarray}
	{\cal M}^{J^P}_{\lambda'\lambda}&=&{\cal M}^{J}_{\lambda'\lambda}+{\eta} {\cal M}^J_{\lambda'-\lambda}
\end{eqnarray}
where $\eta=PP_1P_2(-1)^{J-J_1-J_2}$, with $P$ and $P_{1,2}$ being the parities and $J$ and $J_{1,2}$ being the angular momenta for the system and particle 1 or 2.  It is easy to find that the amplitudes with definite parity have properties such as
\begin{eqnarray}
{\cal M}^{J^P}_{\lambda'-\lambda}&=&\eta {\cal M}^{J\pm}_{\lambda'\lambda},\ \
{\cal M}^{J^P}_{-\lambda'\lambda}=\eta' {\cal M}^{J\pm}_{\lambda'\lambda}.\label{Eq: MM}
\end{eqnarray}
The potential ${\cal V}^{J^P}_{\lambda'\lambda}$ has analogous
relations.

The Bethe-salpeter equation for definite parity can be written as
\begin{eqnarray}
&&{\cal M}^{J^P}_{\lambda'\lambda}={\cal V}^{J^P}_{\lambda'\lambda}+\frac{1}{2}\sum_{\lambda''}{\cal V}^{J^P}_{\lambda'\lambda''}G{\cal M}^{J^P}_{\lambda''\lambda}.\label{Eq: BS_PWA0}
\end{eqnarray}
Please note that there exists a factor $\frac{1}{2}$ on the second term in the right side of the equation.

By using the relation in Eq.~(\ref{Eq: MM}), Eq.~(\ref{Eq: BS_PWA0}) can be rewritten as
\begin{eqnarray}
\hat{\cal M}^{J^P}_{\lambda'\lambda}=\hat{\cal V}^{J^P}_{\lambda'\lambda'}+\sum_{\lambda''\geq0}\hat{\cal V}^{J^P}_{\lambda'\lambda''}G\hat{\cal M}^{J^P}_{\lambda''\lambda},
\end{eqnarray}
where $\lambda$, $\lambda'$ and $\lambda''\geq0$ and $\hat{M}^{J^P}_{\lambda'\lambda}=f_{\lambda'}f_{\lambda} M^{J^P}_{\lambda'\lambda}$, with $f_0=\frac{1}{\sqrt{2}}$ and $f_{\lambda\neq0}=1$.

The sum of the square of the amplitude can be written as a from with definite parity as:
\begin{eqnarray}
\sum_{J,\lambda'\lambda}
	|{\cal M}^{J}_{\lambda'\lambda}|^2
=\sum_{J^P,\lambda'\geq0\lambda\geq0}|\hat{\cal M}^{J^P}_{\lambda'\lambda}|^2.
\end{eqnarray}

By using the normalization of the Wigner $D$ matrix, the integration of the amplitude is
\begin{eqnarray}
\int d\Omega |{\cal M}_{\lambda'\lambda}({\bm p}',{\bm p})|^2=\sum_{J^P,\lambda'\lambda}|\hat{\cal M}^{J^P}_{\lambda'\lambda}({\rm p}',{\rm p})|^2.
\end{eqnarray}
Since there is no interference between the contributions from different partial waves, the
total cross section can also be divided into partial-wave cross sections. Since only the square of the amplitude is related to the physical observables, we omit all hat notation ( $ \hat{} $ )  if not necessary and keep in mind that there is a factor $1/\sqrt{2}$ in the potential also.

\bibliography{../../../reference/Jabref/bibliography}

\end{document}